\documentclass{article}

\usepackage{amssymb,amsfonts,amsmath}
\usepackage{cite,enumerate,float,indentfirst}
\usepackage{color}

\def\be{\begin{eqnarray}}
\def\ee{\end{eqnarray}}
\def\nn{\nonumber}

\def\Tr{{\rm Tr}\,}

\definecolor{red}{rgb}{1,0,0}
\definecolor{orange}{rgb}{1,0.5,0}
\definecolor{violet}{rgb}{0.7,0,1}

\def\cre{\color{red}}

\def\cg{\color{green}}
\def\cb{\color{blue}}

\hoffset= - 1.0in         

\begin{document}

\title{\vspace{-.1cm}{\Large {\bf Correlators in tensor models from character calculus}\vspace{.2cm}}
\author{
{\bf A.Mironov$^{a,b,c}$}\footnote{mironov@lpi.ru; mironov@itep.ru}\ \ and
\ {\bf A.Morozov$^{b,c}$}\thanks{morozov@itep.ru}}
\date{ }
}

\maketitle

\vspace{-5.5cm}

\begin{center}
\hfill FIAN/TD-14/17\\
\hfill IITP/TH-10/17\\
\hfill ITEP/TH-19/17
\end{center}

\vspace{3.3cm}

\begin{center}
$^a$ {\small {\it Lebedev Physics Institute, Moscow 119991, Russia}}\\
$^b$ {\small {\it ITEP, Moscow 117218, Russia}}\\
$^c$ {\small {\it Institute for Information Transmission Problems, Moscow 127994, Russia}}

\end{center}

\vspace{.5cm}

\begin{abstract}
We explain how the calculations of arXiv:1704.08648, which provided
the first evidence for non-trivial structures of Gaussian correlators
in tensor models, are efficiently performed with the help of the
(Hurwitz) character calculus.
This emphasizes a close similarity between technical methods
in matrix and tensor models and supports a hope to understand the
emerging structures in very similar terms.
We claim that the $2m$-fold Gaussian correlators of rank $r$ tensors
are given by $r$-linear combinations of dimensions with the Young diagrams
of size $m$.
The coefficients are made from the characters of the symmetric group $S_m$
and their exact form depends on the choice of the correlator and on the symmetries of the model.
As the simplest application of this new knowledge, we provide simple
expressions for correlators in the Aristotelian tensor model as
tri-linear combinations of dimensions.
\end{abstract}

\bigskip

\bigskip

\section{Introduction}

Emerging interest \cite{tenofirst}-\cite{IMMten2} to tensor models \cite{tensor}
allows one to begin their systematic study.
In the framework of non-linear algebra \cite{NLA}, one does not expect
any essential difference between the tensor and matrix calculi,
and the only difference is that the latter is well developed,
while the former one, not.
Within the systematic approach, the development should proceed in steps,
and the first step is evaluating the Gaussian correlators \cite{Gacor,HZ,MSh}
targeted at finding the underlying structures and their adequate
analytic description.
In \cite{IMMten2}, we demonstrated that the structures are indeed
present and, non-surprisingly, similar to those in matrix models.
To reveal them in full generality and beauty, one, however, needs
to evaluate a lot of quantities, and thus needs an efficient technique
for this.
The goal of this letter is to claim that the most effective calculus
of this kind based on the character expansions \cite{charcalc} and
Hurwitz theory {\it a la} \cite{MMN1}
is directly extended from matrix models case \cite{KR,AMMNhur,MMmamo}
to the tensor case.
A very similar observation is also made in a very recent paper \cite{Di}.
In the present letter, we show how the complicated expressions from \cite{IMMten2}
are drastically simplified by use of the character/Hurwitz calculus.

As explained in big detail in \cite{IMMten2} and \cite{MMmamo},
the simplest for the Gaussian calculus is the rectangular complex matrix model
(RCM) \cite{CM,CMcorr,CMmod}, its tensor liftings are now called {\it rainbow models} \cite{IMMten1}.
The field in this model is the $N_1\times N_2$ matrix $M$,
and the correlators are labeled by Young diagrams $\Lambda=\{m_1\geq m_2\geq \ldots \geq m_{l_\Lambda}> 0\}$:
\be
{\cal O}_\Lambda = \left< \prod_{p=1}^{l_\Lambda} (\Tr M\bar M)^{m_p} \right>
\label{Ocors}
\ee
where $\bar M = M^\dagger$ is Hermitian conjugate, i.e. an $N_2\times N_1$ matrix, while the
averages are defined as the $2N_1N_2$-fold integrals
\be
\Big<\ldots \Big> = \int e^{-\Tr M\bar M} d^2M,
\ \ \ \ \ \ \ \ {\rm with} \ \ \ \ \ \
d^2M = \prod_{a=1}^{N_1}\prod_{\alpha=1}^{N_2} d^2M_{a\alpha}
\ee

In \cite{AMMNhur,IMMten2} and \cite{MMmamo} we described general formulas
for the correlators (\ref{Ocors}).
They possess a simple interpretation in terms of the Hurwitz calculus
and were actually present also in \cite{KR}.
Moreover, their tensorial generalization was also just considered in \cite{Di},
unfortunately, without comparison to the results of \cite{IMMten2}.
Schematically, expressions of this kind for various models look like
(see the main text for the notation)

\bigskip

\begin{tabular}{lcc}
Hermitian matrix model: &  &
$
{\cal O}^{(2m)}_\sigma = \sum_{\gamma \in [2^m]\in S_{2m}} N^{\#(\sigma\circ\gamma)}
$
\\ \\
Rectangular matrix model: &&
$
{\cal O}^{(m)}_\sigma = \sum_{\gamma \in S_m} N_1^\gamma N_2^{\#(\sigma\circ\gamma)}
$
\\ \\
Rainbow tensor model: &&
$
{\cal O}^{(m)}_{\sigma_1\otimes \ldots \otimes\sigma_r} =
\sum_{\gamma \in S_m} \prod_{s=1}^r N_s^{\#(\sigma_s\circ\gamma)}
$
\\ \\
$\ldots$
\end{tabular}

\bigskip

\noindent
The Hurwitz/character calculus allows one to re-express the sums over permutations
in these formulas through multi-linear combinations of dimensions,
like it was done in \cite{MMmamo} in the matrix models case.
As we demonstrate in this letter, the same is possible in the tensor case,
and the formulas are equally simple and powerful.

\section{Description in terms of permutations: RCM
\label{RCM}}

Within the Hurwitz approach, one interprets the Young diagram $R=\{r_p\}$: $r_1\ge r_2\ge\ldots\ge r_{l_R}\ge 0$, $r_i\in\mathbb{Z}$ as that labeling a representation
of the symmetric groups $S_{m}$ with $m = |R| = \sum_{p=1}^{l_R} r_p$ (and also a conjugacy class in the representation)
rather than that of the linear groups (the essence of this calculus is exactly
the explicit description of the Schur-Weyl duality between these two interpretations).
An efficient parametrization of the correlators is possible in terms of permutations
$\sigma \in S_m$,
\be
{\cal O}_\sigma = \left< \prod_{p=1}^{l_\Lambda} (\Tr M\bar M)^{m_p} \right>
= \left<   \prod_{i=1}^m  M_{a_i\alpha_i} {\bar M}^{a_i\alpha_{\sigma(i)}}\right>
\ee
for example,
\be
{\cal O}_{[1]} = \Big<\Tr M\bar M\Big> = \Big< M_{a\alpha} \bar M^{a\alpha}\Big>,
\ \ \ \ \ & \sigma = id =(1)= [1] \in S_1
\nn \\
{\cal O}_{[11]} = \Big<\left(\Tr M\bar M\right)^2\Big> = \Big<
M_{a_1\alpha_1}M_{a_2\alpha_2} \bar M^{a_1\alpha_1}\bar M^{a_2\alpha_2}\Big>,
& \sigma = id = (1)(2)=[11] \in S_2 \nn \\
{\cal O}_{[2]} = \Big<\Tr \left(M\bar M\right)^2\Big> = \Big<
M_{a_1\alpha_1}M_{a_2\alpha_2} \bar M^{a_1\alpha_2}\bar M^{a_2\alpha_1}\Big>,
& \sigma = (12) = [2] \in S_2 \nn \\
\ldots
\ee
More generally, ${\cal O}_{[1^m]} = \Big<\left(\Tr M\bar M\right)^m\Big> $
is associated with $\sigma = id = [1^m] \in S_m$,
while ${\cal O}_{[m]} = \Big<\Tr \left(M\bar M\right)^m\Big> $, with
the longest cycle $\sigma= (12\ldots m)\in S_m$.
In general, the number $l_\Lambda$ of traces is the number of cycles in the associated
$\sigma  \in S_{|\Lambda|}$ (the number of lines in the Young diagram $\Lambda$ describing the conjugacy class of $\sigma$).

It remains to apply the Wick theorem
\be
\left<\prod_{i=1}^m M_{a_i\alpha_i} \bar M^{b_i\beta_{i}}\right>
= \sum_{\gamma \in S_m} \prod_{i=1}^m
\delta_{a_{ i}}^{b_{\gamma(i)}}
\delta_{\alpha_{ i}}^{\beta_{\gamma(i)}}
\label{Wick}
\ee
in order to obtain
\be
\boxed{
{\cal O}_\sigma^{RCM} = {1\over |\sigma|!}\sum_{\gamma \in S_m} N_1^{\#(\gamma)} N_2^{\#(\gamma\circ\sigma)}
}
\label{OasasumMatrix}
\ee
where $\#(\gamma)$ is the number of cycles in the permutation $\gamma$. 
Now one can use the standard identity \cite{Fulton}
\be
p_{\gamma}=\sum_{R\vdash |{\gamma}|}\psi_R({\gamma})\chi_R\{p\}
\ee
Here the sum goes over all Young diagrams with $|{\gamma}|$ boxes, $\chi_R\{p\}$ is the Schur function (the character of the linear group $GL(N)$), $p_{\gamma}=\prod_{i=1}^{l_\gamma}p_{\gamma_i}$ and $\psi_R({\gamma})$ is the character of the symmetric group $S_{|R|}$. Choosing all $p_k=N$, we immediately come to the formula
\be
N^{\#\gamma}=N^{l_\gamma} =
\sum_{R\vdash |\gamma|}   D_R(N)\psi_R(\gamma)
\ee
where $D_R(N)$ are dimensions of representation $R$ of the Lie algebra $GL(N)$, i.e. the value of character at unity.

Further, one can use that, for the symmetric group, $\gamma$ and $\gamma^{-1}$ belong to the same conjugacy class and apply the standard character orthogonality relation (valid for any finite group) \cite{Isaacs}
\be
\sum_\gamma \psi_R(\gamma)\psi_Q(\gamma\circ\sigma)=\sum_\gamma \psi_R(\gamma^{-1})\psi_Q(\gamma\circ\sigma) =
|R|!\
\frac{\psi_R(\sigma)}{d_R}\, \delta_{QR},\ \ \ \ \ \ \ \ d_R={\psi_R(id)\over |R|!}={\psi_R([1^{|R|}])\over |R|!}
\ee
in order to convert the sum (\ref{OasasumMatrix}) into
\be
{\cal O}_\sigma^{RCM} = \sum_{R\vdash m} \frac{D_R(N_1)D_R(N_2)}{d_R}\psi_R(\sigma)
\label{ORCM}
\ee
of \cite{AMMNhur} and \cite{MMmamo}.

\section{Description in terms of permutations: Hermitian matrix model
\label{HM}}

The only difference in this case is that one does not distinguish between $\bar M$
and $M$, which makes the  Wick theorem (\ref{Wick}) a little more involved:
the sum now goes only over permutations $\gamma$ which has $m$ cycles of length $2$,
i.e. $\gamma \in [2^m] \subset S_{2m}$.
This leads to a minor complication in the counterpart of (\ref{ORCM}) \cite{KR,MMmamo}:
\be
{\cal O}_\sigma^{HM} =  \sum_{R\vdash m} \varphi_R([2^m])\cdot D_R(N)\cdot \psi_R(\sigma)
\ee
where $\varphi_R(\mu)$ are again the symmetric group characters,
but with a slightly different normalization, see \cite{MMN1}.

The Hermitian model enumerates equilateral triangulations which can be considered
as ramified coverings of Riemann sphere.
Thus, it is directly related to the ordinary Hurwitz numbers,
Belyi functions, dessins d'enfants and Galois groups,
see \cite{Hermfirst}-\cite{Hermlast} for just some papers on various aspects
 of these relations, and especially \cite{KR} for their connection to formulas
 like (\ref{OasasumMatrix}).

\section{Description in terms of permutations: rainbow tensor models}

Of course, the calculation in sec.\ref{RCM} remains exactly the same
when $M$ is substituted by an arbitrary $N_1\times \ldots \times N_r$ tensor of rank $r$.
The relevant operators are now labeled by sets of $r-1$ permutations
$\vec\sigma=\{\sigma_1,\sigma_2,\ldots,\sigma_r\in S_m\}$,
acting on the corresponding indices in $M$.
The answer for the correlator
is just the obvious generalization of (\ref{OasasumMatrix}):
\be
\boxed{
{\cal O}_{\vec\sigma}^{rainbow}  = \sum_{\gamma \in S_m}
\prod_{s=1}^r N_s^{\#(\gamma\circ\sigma_s)}
}
\label{OasasumTensor}
\ee
and the counterpart of (\ref{ORCM}) is
\be
{\cal O}_{\vec\sigma}^{rainbow} = \sum_{\vec R\vdash m} \prod_{s=1}^r D_{R_s}(N_s)
\psi_{\vec R}(\vec\sigma)
\label{Oten}
\ee
where
\be
\psi_{\vec R}(\vec\sigma) =  \sum_{\gamma\in S_m}
\prod_{s=1}^r\psi_{R_s}(\gamma\circ\sigma_s)
\ee
Unfortunately, there are no known poly-linear counterparts of the orthogonality relation as in the
matrix model case of $r=2$ in order to simplify this formula, however, it is easy to evaluate in every particular example.
Of course, one of the $r$ permutations $\sigma_s$, say, $\sigma_1$ can be absorbed into $\gamma$,
and the correlator depends only on the $r-1$ ``ratios'' $\sigma_1^{-1}\circ\sigma_s$.
Eq.(\ref{Oten}) seems to be in accord with the claim of \cite{Di}.

In the case of non-rainbow models with lower symmetry,
these formulas get a little more complicated,
in exactly the same way as described in sec.\ref{HM}:
some $N$-dependent dimensions $D_{R_s}(N_s)$ are substituted by
symmetric characters, trading the disappearing symmetry parameters $N_s$
for the symmetry-breaking insertions made from $\varphi_R$.

\section{Examples}

\paragraph{Hermitian matrix model:}
\be
\Big< (\Tr X )^2 \Big> = {\cal O}_{ [11]} =
\sum_{\gamma \in [2]\in S_2} N^{\#(\gamma\circ [11])} = N^{\#([2])} = N
\ee

\be
\Big< \Tr (X^2) \Big> = {\cal O}_{  [2]} =
\sum_{\gamma \in [2]\in S_2} N^{\#(\gamma\circ [2])}  = N^{\#([11])} = N^2
\ee

\paragraph{Complex matrix model:}
\be
\Big< (\Tr X\bar X)^2 \Big> = {\cal O}_{id\otimes [11]} =
\sum_{\gamma \in S_2} N_1^{\#(\gamma)}N_2^{\#(\gamma\circ[11])} =
N_1^{\#([11])}N_2^{\#([11])} + N_1^{\#([2])}N_2^{\#([2])} = N_1^2N_2^2+N_1N_2
\ee

\be
\Big< \Tr (X\bar X)^2 \Big> = {\cal O}_{id\otimes [2]} =
\sum_{\gamma \in S_2} N_1^{\#(\gamma)}N_2^{\#(\gamma\circ[2])} = \nn \\
= N_1^{\#([11])}N_2^{\#([2])} + N_1^{\#([2])}N_2^{\#([11])}
= N_1^2N_2+N_1N_2^2 = N_1N_2(N_1+N_2)
\ee

\paragraph{ Aristotelian tensor model \cite{IMMten2}:}
The simplest operators in this rainbow model of the rank-3 tensor $M_{abc}$
with the action $M_{abc}\bar M^{abc}$ are
\be
{\cal K}_m =
M_{a_1b_1c_1} \bar M^{a_1b_2c_2} M_{a_2b_2c_2} \bar M^{a_2b_3c_3} \ldots
M_{a_m b_mc_m}\bar M^{a_mb_1c_1},
\ee
associated with the permutations
\be
\sigma_{{\cal K}_m} = id \otimes (12\ldots m) \otimes (12\ldots m)\in S_m^{\otimes 3},
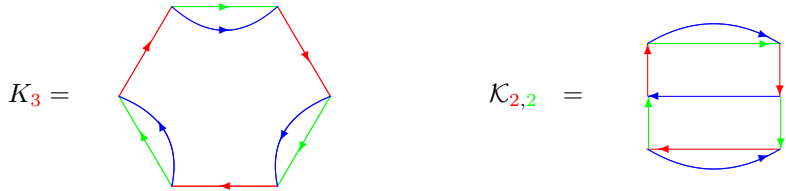
\ee

\begin{picture}(300,110)(-100,-60)

\put(-42,-2){\mbox{$K_{\cre 3}= $}}
{\cre
\qbezier(0,0)(10,17)(20,34)\put(10,17){\vector(1,2){2}}
\qbezier(80,0)(70,17)(60,34)\put(70,17){\vector(1,-2){2}}
\qbezier(20,-34)(40,-34)(60,-34)\put(40,-34){\vector(-1,0){2}}
}
\put(0,0){{\cg
\qbezier(0,0)(10,-17)(20,-34)\put(10,-17){\vector(-1,2){2}}
\qbezier(80,0)(70,-17)(60,-34)\put(70,-17){\vector(-1,-2){2}}
\qbezier(20,34)(40,34)(60,34)\put(40,34){\vector(1,0){2}}
}}
{\cb
\qbezier(0,0)(25,-10)(20,-34)\put(17,-13){\vector(-1,2){2}}
\qbezier(80,0)(55,-10)(60,-34)\put(63,-13){\vector(-1,-2){2}}
\qbezier(20,34)(40,16)(60,34)\put(40,25){\vector(1,0){2}}
}

\put(200,0){
\put(-60,-2){\mbox{${\cal K}_{{\cre 2},{\cg 2}} \ \ = $}}
{\cre
\put(0,0){\vector(0,1){20}}
\put(50,20){\vector(0,-1){20}}
\put(50,-20){\line(-1,0){50}}\put(6,-20){\vector(-1,0){2}}
}
\put(-3,0){
{\cg
\put(50,0){\vector(0,-1){20}}
\put(0,20){\line(1,0){50}}\put(44,20){\vector(1,0){2}}
\put(0,-20){\vector(0,1){20}}
}
}
{\cb
\put(50,0){\vector(-1,0){50}}
\qbezier(0,20)(25,35)(50,20)\put(44,23){\vector(3,-1){2}}
\qbezier(50,-20)(25,-35)(0,-20)\put(44,-23){\vector(3,1){2}}
}
%
}

\end{picture}

\noindent
and ${\cal K}_{m,n}$,
associated with
\be
\sigma_{{\cal K}_{m,n}} =
id \otimes (12\ldots m+n-1) \otimes (12\ldots m)(m+1\ldots m+n-1)\in S_{m+n-1}^{\otimes 3}
\ee

For the average of the simplest ${\cal K}_2$ we have:
\be
\Big<{\cal K}_{\cre 2}\Big> = {\cal O}_{id\otimes [2]\otimes [2]} =
\sum_{\gamma \in S_2} N_1^{\#(\gamma)} N_2^{\#(\gamma\circ[2])} N_3^{\#(\gamma\circ[2])}
= \nn \\
= N_1^{\#([11])}N_2^{\#([2])}N_3^{\#([2])} + N_1^{\#([2])} N_2^{\#([11])} N_3^{\#([11])} =
N_1^2N_2N_3 + N_1N_2^2N_3^2 = N_1N_2N_3\Big(N_1+N_2N_3\Big)
\ee

For the cases of ${\cal K}_3$ and ${\cal K}_{2,2}$ we need multiplication table
in $S_3$ :
\be
\begin{array}{cc|cc|cc}
\gamma &\#(cycles) & \gamma\circ (123) &\#(cycles)& \gamma\circ(12) &\#(cycles)\\
&&&&&\\
\hline
&&&&&\\
id & 3 & (123) & 1 & (12) & 2 \\
(12) & 2 & (13) & 2 & id & 3 \\
(13) & 2 & (23) & 2 & (132) & 1 \\
(23) & 2 & (12) & 2 & (123) & 1 \\
(123) & 1 & (132) & 1 & (23) & 2 \\
(132) & 1 & id & 3 & (13) & 2
\end{array}
\ee
Then
\be
\Big<{\cal K}_{\cre 3}\Big> = {\cal O}_{id\otimes (123)\otimes (123)} =
\sum_{\gamma \in S_3} N_1^{\#(\gamma)} N_2^{\#(\gamma\circ(123))} N_3^{\#(\gamma\circ(123))}
= \nn \\
= N_1^3N_2N_3 + 3N_1^2N_2^2N_3^2+N_1N_2N_3+N_1N_2^3N_3^3 =
N_1N_2N_3\Big(N_2^2N_3^2+3N_1N_2N_3+N_1^2+1\Big)
\ee
\be
\Big<{\cal K}_{{\cre 2},{\cg 2}}\Big> = {\cal O}_{id\otimes (123)\otimes (12)} =
\sum_{\gamma \in S_3} N_1^{\#(\gamma)} N_2^{\#(\gamma\circ(123))} N_3^{\#(\gamma\circ(12))}
= \nn \\
= N_1^3N_2N_3^2 + N_1^2N_2^2N_3^3 + 2N_1^2N_2^2N_3 + N_1N_2N_3^2+N_1N_2^3N_3^2
\ee
All these expressions, indeed, coincide with the answers from \cite{IMMten2}.

\section{On the structure of tensor model correlators}

Eq.(\ref{Oten}) implies that {\bf the correlator in tensor model is actually
a multi-linear combination of dimensions $D_{R_s}(N_s)$ with all $R_s$
of the same size $m$}.
By itself, it is a non-trivial prediction of the Hurwitz formalism,
and it is indeed true, at least for all the examples explicitly evaluated
in \cite{IMMten2} for the Aristotelian model, the simplest non-trivial one within the rainbow class.
For example,
\be
\Big<{\cal K}_{1}\Big> =
D_{[1]}(N_1)D_{[1]}(N_2N_3) =
\boxed{ D_{[1]}\otimes D_{[1]} \otimes D_{[1]}} = N_1N_2N_3
\ee
\be
\Big<{\cal K}_{\cre 2}\Big> =
2\,D_{[2]}(N_1)D_{[2]}(N_2N_3) -2\, D_{[11]}(N_1)D_{[11]}(N_2N_3) = \nn \\
= \boxed{ 2\,\Big(D_{[2]}\otimes D_{[2]}\otimes D_{[2]} + D_{[2]}\otimes D_{[11]}\otimes D_{[11]}
- D_{[11]}\otimes D_{[2]}\otimes D_{[11]} - D_{[11]} \otimes D_{[11]}\otimes D_{[2]}\Big)
} = \nn \\
=\frac{N_1N_2N_3}{4}\Big( (N_1+1)(N_2+1)(N_3+1) + (N_1+1)(N_2-1)(N_3-1) - \nn \\
- (N_1-1)(N_2+1)(N_3-1) - (N_1-1)(N_2-1)(N_3+1)\Big) =
 N_1N_2N_3\Big(N_1 + N_2N_3\Big)
\ee
The first formulas here, bilinear in dimensions are pure matrix model ones,
the new and general, structurally relevant for arbitrary operators
(in particular, more complicated than ${\cal K}_{\cre m}$) are tri-linear expressions
in boxes.
The tensor products mean that the tree dimensions are evaluated at
three different values of $N$.
Note the asymmetry between $N_1$ and $N_2,N_3$ coming from the asymmetry of the operator
${\cal K}_{\cre m}$.
Further,
\be
\Big<{\cal K}_{\cre 3}\Big> =
6\,D_{[3]}(N_1)D_{[3]}(N_2N_3) - 3 \,D_{[21]}(N_1)D_{[21]}(N_2N_3) + 6\,D_{[111]}(N_1)D_{[111]}(N_2N_3)
=\nn\\
= 6\, D_{[3]}\otimes \Big( D_{[3]}\otimes D_{[3]}
+ D_{[21]}\otimes D_{[21]} + D_{[111]}\otimes D_{[111]}\Big) - \nn \\
- 3\,D_{[21]}\otimes \Big(D_{[3]}\otimes D_{[21]}+D_{[21]}\otimes D_{[3]} +
D_{[21]}\otimes D_{[21]} + D_{[21]}\otimes D_{[111]}+D_{[111]}\otimes D_{[21]}\Big) + \nn\\
+ 6\,D_{[111]}\otimes \Big( D_{[3]}\otimes D_{[111]}
+ D_{[21]}\otimes D_{[21]} + D_{[111]}\otimes D_{[3]}\Big)
\ee
\be
\Big<{\cal K}_{{\cre 2},{\cg 2}}\Big> =
\Big(6\, D_{[3]}\otimes D_{[3]}-3\, D_{[21]}\otimes D_{[21]} + 6\,D_{[111]}\otimes D_{[111]}\Big) \otimes D_{[3]} - \nn \\
- \Big(6\, D_{[3]}\otimes D_{[111]}-3\, D_{[21]}\otimes D_{[21]} + 6\,D_{[111]}\otimes D_{[3]}\Big) \otimes D_{[111]}
\ee
Note that there is no matrix model expression (bilinear in dimensions) for
${\cal K}_{{\cre 2},{\cg 2}}$, but the tensor model tri-linear expression exists.

Such formulas allow one to introduce new time variables substituting each dimension
in the tensor product  by a character:
$D_R(N_s) \longrightarrow \chi_R\{p^{(s)}\}$, the original formulas then
arise from these extended quantities at the "classical topological locus" \cite{locus}: $p_k^{(s)} = N_s$.

\section{Conclusion}

In this paper we explained a simple method \cite{Di} to calculate Gaussian averages
in tensor models, which was basically used in evaluation of correlators
in \cite{IMMten2}.
It directly extends the Hurwitz (character) approach to matrix models
\cite{MMN1,KR,AMMNhur,MMmamo} and knot calculus \cite{Sle}.

Technically, it requires knowledge of just two things:

(a) parametrization of operators by permutations and

(b) multiplication table for permutations, actually available
 in MAPLE and Mathematica.

This method proved to be very efficient in matrix model theory \cite{MMN1,KR,AMMNhur,MMmamo}
and it will be equally powerful in application to tensor models.
As emphasized in \cite{IMMten2}, having explicit formulas for correlators
opens a possibility to study recursion relations and genus-like expansions \cite{Gacor,AMM}, Ward identities (Virasoro-like constraints) \cite{MMVir},
the AMM/EO topological recursions \cite{AMM/EO} and many other structures
which are already showing up in the tensor models, very much in the same way
as they did in the matrix model case \cite{UFN3}.

One more aspect of the story is its close relation with the arborescent calculus
\cite{pretzel,arbor,Rama} in knot theory.
Examples in the present text were rather simple, the evaluation of more
sophisticated correlators in tensor models will include the Racah matrices,
not just representation decompositions.

\section*{Acknowledgements}

We are indebted to the referee of our paper \cite{MMmamo} for attracting our attention to the brilliant paper \cite{KR}, which we used in the present text to re-interpret our previous results in a more concise form.

This work was funded by the Russian Science Foundation (Grant No.16-12-10344).


\begin{thebibliography}{12}

\bibitem{tenofirst} E. Witten, arXiv:1610.09758

\bibitem{GuraupostWit} R. Gurau, arXiv:1611.04032; arXiv:1702.04228

\bibitem{KleTar} I. Klebanov, G. Tarnopolsky, Phys.Rev. {\bf D 95} (2017) 046004, arXiv:1611.08915\\
S. Carrozza, A. Tanasa, Letters in Mathematical Physics {\bf 106(11)} (2016) 1531-1559, 1512.06718

\bibitem{Gr} D. Gross, V. Rosenhaus, arXiv:1610.01569; arXiv:1702.08016\\
Ch. Krishnan, S. Sanyal, P.N. Bala Subramanian, arXiv:1612.06330\\
F. Ferrari, 	arXiv:1701.01171\\
V. Bonzom, L. Lionni, A. Tanasa, arXiv:1702.06944\\
M. Beccaria, A.A. Tseytlin, arXiv:1703.04460

\bibitem{SY}  S. Sachdev, Y. Ye, 
Phys.Rev.Lett. {\bf 70} (1993) 3339, cond-mat/9212030\\
J. Polchinski, V. Rosenhaus,
JHEP {\bf 04} (2016) 001, arXiv:1601.06768\\
W. Fu, D. Gaiotto, J. Maldacena, S. Sachdev,
arXiv:1610.08917\\
M. Berkooz, P.Narayan, M. Rozali, J. Simon, 
arXiv:1610.02422

\bibitem{K} A. Kitaev, "A simple model of quantum holography",
http://online.kitp.ucsb.edu/online/entangled15/kitaev/, http:
//online.kitp.ucsb.edu/online/entangled15/kitaev2/. Talks at KITP, April
7, 2015 and May 27, 2015\\
S. Sachdev, 
Phys.Rev. {\bf X5} (2015) 041025, arXiv:1506.05111\\
A. Jevicki, K. Suzuki, J. Yoon, 
arXiv:1603.06246\\
J. Maldacena and D. Stanford, 
arXiv:1604.07818\\
D. Bagrets, A. Altland, A. Kamenev, 
Nucl.Phys. {\bf B911} (2016) 191-205, arXiv:1607.00694\\
A. Jevicki, K. Suzuki, 
arXiv:1608.07567

\bibitem{R1} Z. Bi, C.-M. Jian, Y.-Z. You, K.A. Pawlak, C. Xu, arXiv:1701.07081\\
S.-K. Jian, H. Yao, arXiv:1703.02051\\
S. Carrozza, V .Lahoche, D. Oriti, arXiv:1703.06729\\
Ch. Krishnan, K. Pavan Kumar, S. Sanyal, arXiv:1703.08155\\
M. Casali, P. Cristofori, S. Dartois, L. Grasselli, arXiv:1704.02800\\
Ch. Peng, arXiv:1704.04223\\
S. Das, A. Jevicki, K .Suzuki, arXiv:1704.07208

\bibitem{IMMten1} H.Itoyama, A.Mironov, A.Morozov, Phys.Lett. {\bf B771} (2017) 180-188, arXiv:1703.04983

\bibitem{physlast}
H. Kyono, S. Okumura, K. Yoshida, arXiv:1704.07410

\bibitem{BGRfirst}
R. Gurau, Commun.Math.Phys. {\bf 304} (2011) 69-93, arXiv:0907.2582; Annales Henri Poincare {\bf 11} (2010) 565-584, arXiv:0911.1945;
Class.Quant.Grav. {\bf 27} (2010) 235023, arXiv:1006.0714; Annales Henri Poincare {\bf 13} (2012) 399–423, 1102.5759\\
J.B. Geloun, R. Gurau, V. Rivasseau, Europhys.Lett. {\bf 92} (2010) 60008, arXiv:1008.0354

\bibitem{Gur} R. Gurau, V. Rivasseau, Europhys.Lett. {\bf 95} (2011) 50004, arXiv:1101.4182\\
R. Gurau, J.P. Ryan, SIGMA {\bf 8} (2012) 020, arXiv:1109.4812

\bibitem{Bonz} V. Bonzom, R. Gurau, V. Rivasseau, Phys.Rev. {\bf D85} (2012) 084037, arXiv:1202.3637\\
R. Gurau; V. Rivasseau; S. Gielen, L. Sindoni; J.P. Ryan; V. Bonzom; S. Carrozza; T. Krajewski, R. Toriumi; A. Tanasa; SIGMA {\bf 12} (2016): ''Special Issue on Tensor Models, Formalism and Applications", http://www.emis.de/journals/SIGMA/Tensor\_Models.html

\bibitem{Virtree} V. Bonzom, R. Gurau, A. Riello, V. Rivasseau, 
Nucl.Phys. {\bf B853} (2011) 174-195, arXiv:1105.3122\\
R. Gurau, 
Nucl.Phys. {\bf B852} (2011) 592, arXiv:1105.6072

\bibitem{GurVir} R. Gurau, arXiv:1203.4965\\
V. Bonzom, arXiv:1208.6216

\bibitem{more} V. Bonzom, JHEP {\bf 06} (2013) 062, arXiv:1211.1657\\
V. Bonzom, F. Combes, arXiv:1304.4152\\
V. Bonzom, R. Gurau, J.P. Ryan, A. Tanasa, JHEP {\bf 09} (2014) 05, arXiv:1404.7517\\
R. Gurau, A. Tanasa, D.R. Youmans, Europhys.Lett. {\bf 111} (2015) 21002, arXiv:1505.00586

\bibitem{uncl} A. Tanasa, J.Phys. A: Math.Theor. {\bf 45} (2012) 165401, 1109.0694; SIGMA {\bf 12} (2016) 056, 1512.02087\\
S. Dartois, V. Rivasseau, A. Tanasa, Annales Henri Poincare {\bf 15} (2014) 965-984, 1301.1535

\bibitem{Ram} D. Garner, S. Ramgoolam, Nucl.Phys. {\bf B875} (2013) 244-313, arXiv:1303.3246\\
J.B. Geloun, S. Ramgoolam, arXiv:1307.6490

\bibitem{Cristo} P. Cristofori, E. Fominykh, M. Mulazzani, V. Tarkaev, arXiv:1609.02357

\bibitem{tenolast}
Materials of the 2nd French-Russian Conference on Random Geometry and Physics
(2016), http://www.th.u-psud.fr/RGP16/

\bibitem{IMMten2} H. Itoyama, A. Mironov, A. Morozov, arXiv:1704.08648

\bibitem{tensor} F. David, Nucl.Phys. {\bf B257} (1985) 45\\
V.A. Kazakov, I.K. Kostov, A.A. Migdal,
Phys.Lett. {\bf B157} (1985) 295\\
J. Ambjorn, B. Durhuus, T. Jonsson, Mod.Phys.Lett. {\bf A6} (1991) 1133-1146\\
N. Sasakura, Mod.Phys.Lett. {\bf A6} (1991) 2613\\
P. Ginsparg, hepth/9112013\\
M. Gross, Nucl.Phys.Proc.Suppl. {\bf 25A} (1992) 144-149

\bibitem{NLA} I. Gelfand, M. Kapranov, A. Zelevinsky, {\sl Discriminants, Resultants and Multidimensional Determinants},
Birkhauser, 1994; 
Leningrad Mathematical Journal, {\bf 2:3} (1991) 499-505\\
V.Dolotin, A.Morozov, {\sl Introduction to Non-Linear Algebra}, World Scientific, 2007, hep-th/0609022\\
A. Morozov, Sh. Shakirov, arXiv:0911.5278

\bibitem{Gacor} A. Alexandrov, A. Mironov, A. Morozov,
Int.J.Mod.Phys. {\bf A19} (2004) 4127, hep-th/0310113

\bibitem{HZ} J. Harer, D. Zagier, Invent.Math. {\bf 85} (1986) 457-485\\
C. Itzykson, J.-B. Zuber, Comm.Math.Phys.
{\bf 134} (1990) 197-208\\
S.K. Lando, A.K. Zvonkin, {\sl Embedded graphs}, Max-Plank-Institut f\"ur Mathematik, Preprint 2001 (63)

\bibitem{MSh}
A. Morozov, Sh. Shakirov, JHEP {\bf 0912} (2009) 003, arXiv:0906.0036; arXiv:1007.4100

\bibitem{charcalc}
A. Morozov, Sh. Shakirov,
JHEP {\bf 0904} (2009) 064, arXiv:0902.2627;\\
A. Alexandrov,
arXiv:1005.5715, arXiv:1009.4887;\\
A. Morozov,
Theor.Math.Phys. {\bf 162} (2010) 1-33 (Teor.Mat.Fiz. {\bf 161} (2010) 3-40),
arXiv:0906.3518;\\
A. Balantekin, arXiv:1011.3859

\bibitem{MMN1} A. Mironov, A. Morozov, S. Natanzon,  Theor.Math.Phys. {\bf 166} (2011) 1-22,
arXiv:0904.4227;  Journal of Geometry and Physics {\bf 62} (2012) 148-155,
arXiv:1012.0433; J.Phys. A: Math.Theor. {\bf 45} (2012) 045209,
arXiv:1103.4100

\bibitem{KR} R. de Mello Koch, S. Ramgoolam, arXiv:1002.1634

\bibitem{AMMNhur}
A. Alexandrov, A. Mironov, A. Morozov, S. Natanzon,
JHEP 11 (2014) 080,  arXiv:1405.1395

\bibitem{MMmamo} A. Mironov, A. Morozov, arXiv:1705.00976

\bibitem{Di} P. Diaz and Soo-Jong Rey, arXiv:1706.02667

\bibitem{CM} T. Morris, 
Nucl.Phys. {\bf b356} (1991) 703-728\\
Yu. Makeenko, Pis'ma v ZhETF {\bf 52} (1990) 885-888\\
Yu. Makeenko, A. Marshakov, A. Mironov, A. Morozov, Nucl.Phys. {\bf B356} (1991) 574-628

\bibitem{CMcorr}
C. Kristjansen, J. Plefka, G. W. Semenoff, M. Staudacher, Nucl.Phys. {\bf B643} (2002) 3-30, hep-th/0205033\\
S. Corley, A. Jevicki, S. Ramgoolam, Adv.Theor.Math.Phys. {\bf 5} (2002) 809-839, hep-th/0111222

\bibitem{CMmod}
A. Alexandrov, A. Mironov, A. Morozov, JHEP {\bf 0912} (2009) 053, arXiv:0906.3305

\bibitem{Fulton} W. Fulton, {\sl Young tableaux: with applications to representation theory and geometry},
London Mathematical Society, 1997\\
T. Ceccherini-Silberstein, F. Scarabotti, F. Tolli, {\sl Representation Theory of the
Symmetric Groups}, Cambridge Studies in Advanced Mathematics {\bf 121}, Cambridge University Press, 2010

\bibitem{Isaacs} I.M. Isaacs, {\sl Character Theory of Finite Groups}, (Corrected reprint of the 1976 original, published by Academic Press. ed.), Dover, 1994, ISBN 0-486-68014-2


\bibitem{Hermfirst} G. Belyi, Mathematics of the USSR: Izvestiya, {\bf 14:2} (1980) 247-256\\
A. Grothendieck, {\sl Sketch of a Programme}, Lond. Math. Soc. Lect.
Note Ser. {\bf 242} (1997) 243-283;
{\sl Esquisse d'un Programme}, in: P. Lochak, L. Schneps (eds.), Geometric Galois Action, pp.5-48,
Cambridge University Press, Cambridge (1997)\\
G.B. Shabat, V.A. Voevodsky,
The Grothendieck Festschrift, Birkhauser,
1990, V.III., p.199-227\\
S.K. Lando, A.K. Zvonkin, {\sl Graphs on surfaces and their applications}, Encycl.
of Math. Sciences, {\bf 141}, Springer, 2004

\bibitem{BHM} C. Itzykson, J.B. Zuber, 
Commun. Math. Phys. {\bf 134} (1990) 197\\
T.W. Brown, 
Phys.Rev. {\bf D83}  (2011) 085002, arXiv:1009.0674

\bibitem{LevMor} A. Levin, A. Morozov, 
Phys.Lett. {\bf B243} (1990) 207-214

\bibitem{Gop} R. Gopakumar, arXiv:1104.2386

\bibitem{Hermlast} N. Adrianov, N. Amburg, V. Dremov, Yu. Levitskaya, E. Kreines, Yu .Kochetkov, V. Nasretdinova, G. Shabat, arXiv:0710.2658

\bibitem{locus} P. Dunin-Barkowski, A. Mironov, A. Morozov, A. Sleptsov, A. Smirnov,
  JHEP {\bf 03} (2013) 021,  arXiv:1106.4305\\
A. Mironov, A. Morozov, An. Morozov, in: {\sl Strings, Gauge Fields, and the Geometry Behind: The Legacy of Maximilian Kreuzer}, edited by A. Rebhan, L. Katzarkov, J. Knapp, R. Rashkov, E. Scheidegger (World Scietific Publishins Co.Pte.Ltd. 2013) pp.101-118, arXiv:1112.5754;   JHEP {\bf 03} (2012) 034,  arXiv:1112.2654

\bibitem{Sle} A. Mironov, A. Morozov, A.Sleptsov, 
Theor.Math.Phys. {\bf 177} (2013) 179-221, arXiv:1303.1015;
The European Physical Journal, {\bf C73} (2013) 2492, arXiv:1304.7499\\
A. Mironov, A. Morozov, A. Sleptsov, A. Smirnov, Nucl.Phys. {\bf B889} (2014) 757-777, arXiv:1310.7622\\
A. Sleptsov, 
Int.J.Mod.Phys. {\bf A29} (2014) 1430063

\bibitem{AMM} A. Alexandrov, A. Mironov, A. Morozov, Int.J.Mod.Phys. {\bf A21} (2006) 2481-2518,
hep-th/0412099; Fortsch.Phys. {\bf 53} (2005) 512-521, hep-th/0412205\\
A. Mironov, A. Morozov, arXiv:1701.03057

\bibitem{MMVir} F. David, Mod.Phys.Lett. {\bf A5} (1990) 1019\\
A. Mironov, A. Morozov, Phys.Lett. {\bf B252} (1990) 47-52\\
J. Ambj{\o}rn, Yu. Makeenko, Mod.Phys.Lett. {\bf A5} (1990) 1753\\
H. Itoyama, Y. Matsuo, Phys.Lett. {\bf 255B} (1991) 20

\bibitem{AMM/EO} A. Alexandrov, A. Mironov, A. Morozov,
Physica {\bf D235} (2007) 126-167, hep-th/0608228; JHEP {\bf 12} (2009) 053, arXiv:0906.3305\\
 B. Eynard, N. Orantin, Commun. Number Theory Phys. {\bf 1} (2007) 347-452, math-ph/0702045\\
N. Orantin,
arXiv:0808.0635

\bibitem{UFN3} A. Morozov,
Phys.Usp.(UFN) {\bf 37} (1994) 1;
hep-th/9502091; hep-th/0502010\\
A. Mironov, Int.J.Mod.Phys. {\bf A9} (1994) 4355; Phys.Part.Nucl.
{\bf 33} (2002) 537; hep-th/9409190

\bibitem{pretzel}
D. Galakhov, D. Melnikov, A. Mironov, A. Morozov, A. Sleptsov, Phys.Lett. {\bf B743} (2015) 71-74, arXiv:1412.2616\\
A. Mironov, A. Morozov, A. Sleptsov, JHEP {\bf 07} (2015) 069,  arXiv:1412.8432

\bibitem{arbor} P. Ramadevi, T.R. Govindarajan, R.K. Kaul, Mod.Phys.Lett. {\bf A9} (1994) 3205-3218, hep-th/9401095\\
S. Nawata, P. Ramadevi, Zodinmawia,
J.Knot Theory and Its Ramifications {\bf 22} (2013) 13, arXiv:1302.5144\\
Zodinmawia's PhD thesis, 2014\\
S. Nawata, P. Ramadevi, Vivek Kumar Singh,  arXiv:1504.00364

\bibitem{Rama} A. Mironov, A. Morozov, An. Morozov, P. Ramadevi, Vivek Kumar Singh,
JHEP {\bf 1507} (2015) 109,  arXiv:1504.00371\\
A.Mironov, A.Morozov, Nucl.Phys. {\bf B899} (2015) 395-413,
 arXiv:1506.00339\\
A. Mironov, A. Morozov, An. Morozov, P. Ramadevi, Vivek Kumar Singh, A. Sleptsov, J.Phys. {\bf A50} (2017) 085201, arXiv:1601.04199

\end{thebibliography}
\end{document}